\documentclass[11pt,draftcls,onecolumn,peerreviewca]{IEEEtran}

\usepackage[cmex10]{amsmath}
\usepackage{graphicx}
\usepackage{amssymb}
\usepackage{pseudocode}
\usepackage{cite}

\newtheorem{proposition}{Proposition}

\newtheorem{lemma}{Lemma}

\allowdisplaybreaks[3]

\begin{document}

\title{Robust Secure Transmission in MISO Channels Based on Worst-Case Optimization}

\author{Jing~Huang,~\IEEEmembership{Student~Member,~IEEE} and
				A.~Lee~Swindlehurst,~\IEEEmembership{Fellow,~IEEE}
\thanks{The authors are with the Dept. of Electrical Engineering \& Computer Science, University of California, Irvine, CA 92697-2625, USA. (Phone: 1-949-824-1818, Fax: 1-949-824-3779, Email: \{jing.huang; swindle\}@uci.edu). }
\thanks{This work was supported by the U.S. Army Research Office under the Multi-University Research
Initiative (MURI) grant W911NF-07-1-0318.}
}

\begin{titlepage}
\maketitle
\begin{abstract}
This paper studies robust transmission schemes for multiple-input single-output (MISO) wiretap channels. Both the cases of direct transmission and cooperative jamming with a helper are investigated with imperfect channel state information (CSI) for the eavesdropper links. Robust transmit covariance matrices are obtained based on worst-case secrecy rate maximization, under both individual and global power constraints. For the case of an individual power constraint, we show that the non-convex maximin optimization problem can be transformed into a quasiconvex problem that can be efficiently solved with existing methods. For a global power constraint, the joint optimization of the transmit covariance matrices and power allocation between the source and the helper is studied via geometric programming. We also study the robust wiretap transmission problem for the case with a quality-of-service constraint at the legitimate receiver. Numerical results show the advantage of the proposed robust design. In particular, for the global power constraint scenario, although cooperative jamming is not necessary for optimal transmission with perfect eavesdropper's CSI, we show that robust jamming support can increase the worst-case secrecy rate and lower the signal to interference-plus-noise ratio at Eve in the presence of channel mismatches between the transmitters and the eavesdropper.
\end{abstract}

\end{titlepage}

\section{Introduction}
Secure transmission in wireless networks is required in many applications. Traditionally, security is considered as an issue addressed above the physical (PHY) layer, and conventional approaches for ensuring confidentially are usually based on cryptographic methods. However, the broadcast nature of wireless transmission and the dynamic topology of mobile networks may introduce significant challenges to secret key transmission and management \cite{Debbah_Mobile08,Liang_Information09}. Therefore, there has recently been considerable interest from an information-theoretic perspective in the use of physical layer mechanisms to improve the security of wireless transmissions.

The theoretical basis of this area was initiated by Wyner, who introduced and studied the wiretap channel where the eavesdropper's received signal is a degraded version of the legitimate receiver's signal \cite{Wyner_wire-tap75}. The secrecy capacity was defined as the upper bound of all achievable secrecy rates, which guarantee that private messages can be reliably transmitted to the receiver and kept perfectly secret from the eavesdropper. Csisz\'{a}r and K\"{o}rner studied a more general non-degraded wiretap channel and considered transmission of secret messages over broadcast channels \cite{Csiszar_Broadcast78}. Recently, considerable research has investigated secrecy in wiretap channels with multiple antennas \cite{Goel_Guaranteeing08, Parada_Secrecy05, Khisti_Secure10, Liu_Note09,Shafiee_Achievable07,Liu_MMSE09, Oggier_secrecy08,Hero_Secure03,Khisti_Gaussian07}. The secrecy capacity of the multiple-input multiple-output (MIMO) wiretap channel has been fully characterized in \cite{Liu_Note09} and \cite{Oggier_secrecy08}. In particular, for multiple-input single-output (MISO) wiretap cannels, the optimal transmit covariance matrix was found to be single-stream beamforming obtained via a closed-form solution \cite{Li_Secret07,Khisti_Secure10a}. While research in this area usually assumes global channel state information (CSI) is available at the transmitter, some other work has considered the case where only partial information of the eavesdropper CSI (ECSI) is available. The optimal transmit covariance matrix that achieves the ergodic secrecy capacity for the MISO wiretap channel was studied in \cite{Li_Ergodic10}, where only statistical ECSI is assumed to be available. In \cite{Wolf_Maximization10} and \cite{Zhang_Robust09}, the problem was investigated from the perspective of maximizing the worst-case secrecy rate, where \cite{Wolf_Maximization10} considered the case that the eavesdropper's channel matrix is trace-bounded but otherwise unknown, and \cite{Zhang_Robust09} assumed that the ECSI channel mismatch is norm-bounded.

With the additional degrees of freedom in multi-antenna or multi-node systems, many papers have considered improving the secrecy rate through the use of artificial interference \cite{Goel_Guaranteeing08,Swindlehurst_Fixed09,Khisti_Gaussian07,Mukherjee_Robust11}. In these schemes, in addition to the information signals, part of the transmit power is allocated to jamming signals that selectively degrade the eavesdropper's channel while maintaining little interference to legitimate users. Some recent work has also considered using friendly helpers to provide jamming signals to confuse the eavesdropper \cite{Tekin_General08,Lai_Relay--Eavesdropper08,Wang_Cooperative09,Dong_Improving10,Li_Optimal10,Huang_Secure10,Zheng_Optimal11,Wolf_zero10}. This approach is often referred to as cooperative jamming. In \cite{Tekin_General08}, a cooperative jamming scheme is proposed for improving the achievable secrecy sum-rate for general Gaussian multiple access and two-way relay wiretap channels. The optimal transmit weights for multiple single-antenna helpers were studied in \cite{Dong_Improving10,Li_Optimal10}, where a global power constraint was imposed. A similar case with individual power constraints was studied in \cite{Zheng_Optimal11}. In \cite{Wolf_zero10}, the optimal beamforming strategy for a cooperative jammer was studied for the MISO case under a zero-forcing constraint that nulls the interference at the legitimate receiver. However, most of the previous work on cooperative jamming assumes perfect global channel state information, including CSI for the eavesdropper. This motivates us to investigate the case when the transmitters (including helpers) have only imperfect ECSI.

In this paper, we study robust transmit precoder design for MISO wiretap channels with and without a helper. We assume that perfect CSI for the links to the legitimate user is available at both transmitters, while for the eavesdropper links there exist channel mismatches that are norm-bounded by some known constants \cite{Zhang_Robust09,Vorobyov_Robust03,Wang_Worst-Case09}. Following \cite{Khisti_Secure10a,Oggier_secrecy08,Shafiee_Achievable07}, Gaussian inputs are assumed in the paper. We focus on obtaining robust transmit covariance matrices for 1) direct transmission (DT) without a helper and 2) cooperative jamming (CJ) schemes with a helper (friendly jammer), based on maximizing the worst-case secrecy rate. Note that our work is different from \cite{Zhang_Robust09} in that we consider robust transmission for not only information signals but also jamming signals. We begin by studying the optimization problem under an individual power constraint, and then focus on the more complicated case with a global power constraint. For the individual power constraint case, we show that the non-convex maximin problem of maximizing the worst-case secrecy rate can be converted into a quasiconvex problem that can be efficiently solved via existing methods. For the global power constraint case, we obtain the jointly optimal transmit covariance matrices and power allocation between the source and the helper via geometric programming. In addition, following \cite{Swindlehurst_Fixed09,Mukherjee_Fixed-rate09,Mukherjee_Robust11}, we also consider the scenario where there is a quality-of-service (QoS) constraint at the legitimate node. In this case, we show that the introduction of the QoS constraint simplifies the optimization of the covariance and the power allocation.

The organization of the paper is as follows. Section \ref{sec:sm} describes the system model considered throughout the paper. In Section \ref{sec:rdt}, robust design of the transmit covariance matrix is studied for the direct transmission case. The robust cooperative jamming scheme is then investigated in Section \ref{sec:rcj}, where both individual and global power constraints are considered. Section \ref{sec:qos} studies the case where a QoS constraint is required at the legitimate destination. The performance of the proposed robust transmission approaches are studied using several simulation examples in Section \ref{sec:nr}, and conclusions are drawn in Section \ref{sec:con}.

The following notation is used in the paper: $\mathbb{E} \{\cdot\}$ denotes expectation, $(\cdot)^H$ the Hermitian transpose, $||\cdot||$ the Euclidean norm, $(\cdot)^\dag$ the pseudo-inverse, $\textrm{tr}(\cdot)$ is the trace operator, $\mathcal{R}(\cdot)$ represents the range space of a matrix, and $\mathbf{I}$ is an identity matrix of appropriate dimension. 
\section{System Model} \label{sec:sm}
We consider a MISO communication system with a source node
(Alice), a helper (Helper), a destination (Bob), and an eavesdropper
(Eve). The number of antennas possessed by Alice and the Helper are denoted by $N_a$ and $N_h$, respectively, while both Bob and Eve are single-antenna nodes. In this model, Alice sends private messages to Bob in the presence of Eve, who is able to eavesdrop on the link between Alice and Bob. The Helper can choose to be silent or to transmit artificial interference signals to confuse Eve. Both cases will be considered in the paper, and we refer to the former case as direct transmission (DT) and the latter as cooperative jamming (CJ). We assume that Alice and the Helper have perfect CSI for their links to Bob, but they have only imperfect CSI for their channels to Eve. We will consider cases with either individual or global power constraints.

\subsection{Direct Transmission} \label{sec:dt}
When there is no support from the Helper, the received signals at Bob and Eve are given by
\begin{subequations} \label{eq:signojam}
\begin{align}
y_b &= \mathbf{h}_{b} \mathbf{x} + n_b \\
y_e &= \mathbf{h}_{e} \mathbf{x} + n_e
\end{align}
\end{subequations}
where $\mathbf{x}$ is the signal vector transmitted by Alice, the covariance matrix of $\mathbf{x}$ is denoted by $\mathbf{Q}_x = \mathbb{E}\{\mathbf{x} \mathbf{x}^H \}$, $\textrm{tr}(\mathbf{Q}_x) \le P_S$ where $P_S$ is the transmit power constraint on Alice, and $\{ \mathbf{h}_{b}, \mathbf{h}_{e} \}$ are the $1 \times N_a$ channel vectors for Bob and Eve, respectively. The terms $n_b$ and $n_e$ represent naturally occurring noise at Bob and Eve, and  we assume that $n_b$ and $n_e$ are zero-mean circular complex Gaussian with variance $\sigma_b^2$ and $\sigma_e^2$. We will assume without loss of generality that $\sigma_b^2=\sigma_e^2=\sigma^2$.

\subsection{Cooperative Jamming}
For the case where the Helper joins the network by transmitting an i.i.d. Gaussian interference signal $\mathbf{z}$, Bob and Eve then receive
\begin{subequations} \label{eq:sigjam}
\begin{align}
y_b &= \mathbf{h}_{b} \mathbf{x} + \mathbf{g}_{b} \mathbf{z} + n_b \\
y_e &= \mathbf{h}_{e} \mathbf{x} + \mathbf{g}_{e} \mathbf{z} + n_e
\end{align}
\end{subequations}
where we denote $\mathbf{Q}_z = \mathbb{E}\{\mathbf{z} \mathbf{z}^H \}$ and $\textrm{tr}(\mathbf{Q}_z) \le P_J$. The cooperative jamming optimization problems that we consider in the paper will be subject to either an individual power constraint $\textrm{tr}(\mathbf{Q}_x) \le P_S$, $\textrm{tr}(\mathbf{Q}_z) \le P_J$, or to a global power constraint $\textrm{tr}(\mathbf{Q}_z) + \textrm{tr}(\mathbf{Q}_z) \le P$.

\subsection{Channel Mismatch}
For the channels between the transmitters and Eve, only estimates $\tilde{\mathbf{h}}_e$ and $\tilde{\mathbf{g}}_e$ are available at Alice and the Helper, respectively. We define the channel error vectors as
\begin{subequations} \label{eq:mismt}
\begin{align}
\mathbf{e}_h &= \mathbf{h}_e - \tilde{\mathbf{h}}_e \\
\mathbf{e}_g &= \mathbf{g}_e - \tilde{\mathbf{g}}_e,
\end{align}
\end{subequations}
and we assume that the channel mismatches lie in the bounded sets
$\mathcal{E}_h = \{\mathbf{e}_h: ||\mathbf{e}_h||^2  \le \epsilon_h^2\}$ and $\mathcal{E}_g = \{\mathbf{e}_g: ||\mathbf{e}_g||^2 \le \epsilon_g^2\}$, where $\epsilon_h$ and $\epsilon_g$ are known constants. All the optimization problems in the paper are based on the $\mathbf{e}_h^* \in \mathcal{E}_h$ and  $\mathbf{e}_g^* \in \mathcal{E}_g$ that give the worst performance. 
\section{Robust Direct Transmission} \label{sec:rdt}
In this section, we consider the scenario where there is no jamming support from the Helper. According to the signal model \eqref{eq:signojam} and \eqref{eq:mismt}, the secrecy rate is \cite{Shafiee_Achievable07}
\begin{equation} \label{eq:rdtsr}
R_s = \log_2 \left( 1+\frac{1}{\sigma^2} \mathbf{h}_b \mathbf{Q}_x \mathbf{h}_b^H \right)
- \log_2 \left( 1+\frac{1}{\sigma^2} (\tilde{\mathbf{h}}_e + \mathbf{e}_h) \mathbf{Q}_x (\tilde{\mathbf{h}}_e^H +\mathbf{e}_h) \right).
\end{equation}
A power constraint is imposed such that $\mathbf{Q}_x \in \mathcal{Q}_x = \{\mathbf{Q}_x: \mathbf{Q}_x \succeq 0, {\textrm{tr}} (\mathbf{Q}_x) \le P_S\}$. For the case where perfect ECSI is available, the optimal $\mathbf{Q}_x$ has been found to be unit-rank and the corresponding beamformer is the generalized eigenvector of the matrix pencil $(\sigma^2 \mathbf{I} + P_S \mathbf{h}_{b}^H \mathbf{h}_{b}, \sigma^2 \mathbf{I} + P_S \mathbf{h}_{e}^H \mathbf{h}_{e})$ corresponding to the largest generalized eigenvalue \cite{Shafiee_Achievable07,Khisti_Secure10a}.

We consider the case where Alice does not have perfect knowledge of the channel to Eve, but only an estimate $\tilde{\mathbf{h}}_e$.
We focus on optimizing the worst-case performance, where we maximize the secrecy rate for the worst channel mismatch $\mathbf{e}_h$ in the bounded set $\mathcal{E}_h$. Therefore, the optimization problem \eqref{eq:rdtsr} becomes
\begin{equation} \label{eq:maximin1}
\max_{\mathbf{Q}_x \in \mathcal{Q}_x} \min_{\mathbf{e}_h \in \mathcal{E}_h}
\frac{\sigma^2+\mathbf{h}_b \mathbf{Q}_x \mathbf{h}_b^H}
{\sigma^2+ (\tilde{\mathbf{h}}_e+\mathbf{e}_h) \mathbf{Q}_x (\tilde{\mathbf{h}}_e+\mathbf{e}_h)^H}.
\end{equation}
The difficulty in solving this problem comes from the inner minimization over $\mathbf{e}_h$. As will be discussed later, the minimization is actually a non-convex problem. However, we will show that through a proper transformation, problem \eqref{eq:maximin1} can be converted to a solvable quasiconvex optimization problem.
\begin{proposition} \label{th:maxqx}
Problem \eqref{eq:maximin1} is equivalent to the following problem
\begin{subequations} \label{eq:lfp}
\begin{align}
\min_{\mathbf{Q}_x, \mu, \mathbf{\Psi}}& \quad
\frac{\sigma^2+\mu \epsilon_h^2 + \textrm{tr} [(\mathbf{Q}_x+\mathbf{\Psi}) \tilde{\mathbf{h}}_e^H \tilde{\mathbf{h}}_e]} {\sigma^2+\textrm{tr}( \mathbf{Q}_x \mathbf{h}_b^H \mathbf{h}_b)} \\
{\textrm{s.t.}}& \quad
\left[
\begin{array}{cc}
\mathbf{\Psi} & \mathbf{Q}_x \\
\mathbf{Q}_x & \mu \mathbf{I}_{N_a} -\mathbf{Q}_x
\end{array}
\right ] \succeq 0 \\
					 & \quad   \textrm{tr}(\mathbf{Q}_x) \le P_S \\
					 & \quad   \mathbf{Q}_x \succeq 0, \mu \ge 0.
\end{align}
\end{subequations}
\end{proposition}
\begin{IEEEproof}
The maximin problem in \eqref{eq:maximin1} can be transformed to
\begin{subequations}
\begin{align}
\max_{\mathbf{Q}_x \in \mathcal{Q}_x, v}& \quad \frac{\sigma^2+\mathbf{h}_b \mathbf{Q}_x \mathbf{h}_b^H}{v} \\
{\textrm{s.t.}}& \quad \sigma^2+ (\tilde{\mathbf{h}}_e+\mathbf{e}_h) \mathbf{Q}_x (\tilde{\mathbf{h}}_e+\mathbf{e}_h)^H \le v \\
					 & \quad \mathbf{e}_h \mathbf{e}_h^H \le \epsilon_h^2,
\end{align}
\end{subequations}
where the constraints can also be expressed as
\begin{align}
&-\mathbf{e}_h \mathbf{Q}_x \mathbf{e}_h^H - 2 {\textrm{Re}}(\tilde{\mathbf{h}}_e \mathbf{Q}_x \mathbf{e}_h^H)
- \tilde{\mathbf{h}}_e \mathbf{Q}_x \tilde{\mathbf{h}}_e^H - \sigma^2 + v \ge 0  \label{eq:sproc1}\\
&- \mathbf{e}_h \mathbf{e}_h^H + \epsilon_h^2 \ge 0.
\end{align}
Using the $\mathcal{S}$-procedure \cite{Boyd_Convex04}, we know that there exists an $\mathbf{e}_h \in \mathbb{C}^{N_a}$ satisfying both the above inequalities if and only if there exists a $\mu \ge 0$ such that
\begin{equation} \label{eq:lhs}
\left[
\begin{array}{cc}
\mu \mathbf{I}_{N_a} - \mathbf{Q}_x & -\mathbf{Q}_x \tilde{\mathbf{h}}_e^H \\
-\tilde{\mathbf{h}}_e \mathbf{Q}_x & -\tilde{\mathbf{h}}_e \mathbf{Q}_x \tilde{\mathbf{h}}_e^H-\sigma^2-\mu \epsilon_h^2 + v
\end{array}
\right] \succeq 0.
\end{equation}
Then we can use the property of the generalized Schur complement \cite{Carlson_Generalization74} and rewrite \eqref{eq:lhs} as
\begin{equation} \label{eq:eqv}
\sigma^2+\mu \epsilon_h^2 + \tilde{\mathbf{h}}_e \mathbf{Q}_x \tilde{\mathbf{h}}_e^H + \tilde{\mathbf{h}}_e \mathbf{Q}_x
(\mu \mathbf{I}_{N_a} - \mathbf{Q}_x)^\dag \mathbf{Q}_x \tilde{\mathbf{h}}_e^H \le v.
\end{equation}
Therefore, the maximin problem in \eqref{eq:maximin1} becomes
\begin{equation} \label{eq:sproc2}
\max_{\mathbf{Q}_x \in \mathcal{Q}_x, \mu \ge 0} \quad
\frac{\sigma^2+\mathbf{h}_b \mathbf{Q}_x \mathbf{h}_b^H}
{\sigma^2+\mu \epsilon_h^2 + \tilde{\mathbf{h}}_e \mathbf{Q}_x \tilde{\mathbf{h}}_e^H + \tilde{\mathbf{h}}_e \mathbf{Q}_x
(\mu \mathbf{I}_{N_a} - \mathbf{Q}_x)^\dag \mathbf{Q}_x \tilde{\mathbf{h}}_e^H}
\end{equation}
which is equivalent to
\begin{align}
\max_{\mathbf{Q}_x \in \mathcal{Q}_x, \mu \ge 0, \mathbf{\Psi}}& \quad
\frac{\sigma^2+\mathbf{h}_b \mathbf{Q}_x \mathbf{h}_b^H}
{\sigma^2+\mu \epsilon_h^2 + \tilde{\mathbf{h}}_e \mathbf{Q}_x \tilde{\mathbf{h}}_e^H + \tilde{\mathbf{h}}_e \mathbf{\Psi} \tilde{\mathbf{h}}_e^H} \\
{\textrm{s.t.}}& \quad   \mathbf{Q}_x
(\mu \mathbf{I}_{N_a} - \mathbf{Q}_x)^\dag \mathbf{Q}_x \preceq \mathbf{\Psi}. \label{eq:schurconv}
\end{align}
Next, we use the Schur complement to convert \eqref{eq:schurconv} into a linear matrix inequality (LMI), and the maximization problem is then given by
\begin{subequations}
\begin{align}
\min_{\mathbf{Q}_x \in \mathcal{Q}_x, \mu \ge 0, \mathbf{\Psi}}& \quad
\frac{\sigma^2+\mu \epsilon_h^2 + \textrm{tr} [(\mathbf{Q}_x+\mathbf{\Psi}) \tilde{\mathbf{h}}_e^H \tilde{\mathbf{h}}_e]} {\sigma^2+\textrm{tr}( \mathbf{Q}_x \mathbf{h}_b^H \mathbf{h}_b)} \\
{\textrm{s.t.}}& \quad
\left[
\begin{array}{cc}
\mathbf{\Psi} & \mathbf{Q}_x \\
\mathbf{Q}_x & \mu \mathbf{I}_{N_a} -\mathbf{Q}_x
\end{array}
\right ] \succeq 0,
\end{align}
\end{subequations}
which completes the proof.
\end{IEEEproof}

 Problem \eqref{eq:lfp} consists of a linear fractional objective function, which is thus quasiconvex, with a set of LMI constraints. Therefore,  we can solve this problem efficiently via the bisection method \cite{Boyd_Convex04}. We first form the epigraph problem for \eqref{eq:lfp} as
\begin{subequations} \label{eq:epi}
\begin{align}
\min_{\mathbf{Q}_x, \mu, \mathbf{\Psi}}& \quad t \\
{\textrm{s.t.}}
					 & \quad \sigma^2+\mu \epsilon_h^2 + {\textrm{tr}} [(\mathbf{Q}_x+\mathbf{\Psi}) \tilde{\mathbf{h}}_e^H \tilde{\mathbf{h}}_e] \le  t~[\sigma^2 + {\textrm{tr}}( \mathbf{Q}_x \mathbf{h}_b^H \mathbf{h}_b) ] \\
					 & \quad
\left[
\begin{array}{cc}
\mathbf{\Psi} & \mathbf{Q}_x \\
\mathbf{Q}_x & \mu \mathbf{I}_{N_a} -\mathbf{Q}_x
\end{array}
\right ] \succeq 0 \\
					 & \quad   \textrm{tr}(\mathbf{Q}_x) \le P_S, \mathbf{Q}_x \succeq 0, \mu \ge 0,
\end{align}
\end{subequations}
and the corresponding feasibility problem is then given by
\begin{subequations} \label{eq:feap}
\begin{align}
{\textrm{Find}}& \quad \mathbf{Q}_x, \mu, \mathbf{\Psi} \\
{\textrm{s.t.}}
					 & \quad \sigma^2+\mu \epsilon_h^2 + {\textrm{tr}} [(\mathbf{Q}_x+\mathbf{\Psi}) \tilde{\mathbf{h}}_e^H \tilde{\mathbf{h}}_e] \le  t~[\sigma^2 + {\textrm{tr}}( \mathbf{Q}_x \mathbf{h}_b^H \mathbf{h}_b) ] \\
					 & \quad
\left[
\begin{array}{cc}
\mathbf{\Psi} & \mathbf{Q}_x \\
\mathbf{Q}_x & \mu \mathbf{I}_{N_a} -\mathbf{Q}_x
\end{array}
\right ] \succeq 0 \\
					 & \quad   \textrm{tr}(\mathbf{Q}_x) \le P_S, \mathbf{Q}_x \succeq 0, \mu \ge 0.
\end{align}
\end{subequations}
Then we can use the following bisection method by solving the convex feasibility problem at each step until the interval that contains the optimal value converges:

\begin{pseudocode}[framebox]{Bisection method for obtaining $\mathbf{Q}_x$}{}
\INIT \TX{set an interval } [l,u] \TX{ such that the optimal value } t^* \in [l,u], \TX{ and define a tolerance } \delta > 0. \\
\WHILE u-l>\delta \\
\DO
\BEGIN
\LET t=(l+u)/2. \\
\TX{Solve the feasibility problem in \eqref{eq:feap}}.\\
\IF \TX{\eqref{eq:feap} is feasible,} \THEN u=t; \ELSE l=t.
\END
\end{pseudocode}\\
For problem \eqref{eq:lfp}, the initial lower bound $l$ can be chosen as any infeasible value, for example $l=\frac{\sigma^2}{\sigma^2+P_S ||\mathbf{h}_b||^2}$, which is a lower bound since $\textrm{tr}(\mathbf{Q}_x \mathbf{h}_b^H \mathbf{h}_b) \le ||\mathbf{Q}_x||_2 \cdot ||\mathbf{h}_b||^2 \le P_S ||\mathbf{h}_b||^2$. The upper bound $u$ can be any feasible value, for instance, we can set $u=\frac{\sigma^2+\mu_0 \epsilon_h^2 + \textrm{tr} [(\mathbf{Q}_0+\mathbf{\Psi}_0) \tilde{\mathbf{h}}_e^H \tilde{\mathbf{h}}_e]} {\sigma^2+\textrm{tr}( \mathbf{Q}_0 \mathbf{h}_b^H \mathbf{h}_b)}$, where $\mu_0=P_S$, $\mathbf{Q}_0=\frac{P_S}{N_a} \mathbf{I}$, and $\mathbf{\Psi}_0 = \mathbf{Q}_0 (\mu_0 \mathbf{I} - \mathbf{Q}_0)^\dag \mathbf{Q}_0$.

Note that the solution for the optimal covariance $\mathbf{Q}_x^*$ obtained from Proposition \ref{th:maxqx} is already based on a hidden worst-case channel mismatch $\mathbf{e}_h^*$. Next, we will explicitly express $\mathbf{e}_h^*$ under the norm-bounded constraint, which will be useful for the joint optimization in Section \ref{sec:globalp}. The problem is formulated as
\begin{subequations} \label{eq:maxeh}
\begin{align} \label{eq:wstdel1}
\max_{\mathbf{e}_h}& \quad (\tilde{\mathbf{h}}_e+\mathbf{e}_h) \mathbf{Q}_x^* (\tilde{\mathbf{h}}_e+\mathbf{e}_h)^H \\
\textrm{s.t.}& \quad ||\mathbf{e}_h|| \le \epsilon_h.
\end{align}
\end{subequations}
This is a non-convex problem since we want to maximize a convex function. However, we can still obtain the global optimum by solving its dual problem, as explained in the following proposition.
\begin{proposition} \label{th:weh}
The worst-case channel mismatch for problem \eqref{eq:maxeh} is given by $\mathbf{e}_h^* = \tilde{\mathbf{h}}_e \mathbf{Q}_x^* (\lambda \mathbf{I} - \mathbf{Q}_x^*)^\dag$, where $\lambda$ is the solution of the following problem
\begin{subequations} \label{eq:trustr1}
\begin{align}
\max_{\lambda \ge 0, \gamma}& \quad \gamma \\
{\textrm{s.t.}}& \quad
\left[
\begin{array}{cc}
\lambda \mathbf{I} - \mathbf{Q}_x^* & \mathbf{Q}_x^* \tilde{\mathbf{h}}_e^H \\
\tilde{\mathbf{h}}_e \mathbf{Q}_x^* & - \tilde{\mathbf{h}}_e \mathbf{Q}_x^* \tilde{\mathbf{h}}_e^H - \lambda \epsilon_h^2 - \gamma
\end{array}
\right ] \succeq 0.
\end{align}
\end{subequations}
\end{proposition}
\begin{IEEEproof}
See Appendix \ref{sec:wehpf}.
\end{IEEEproof}

Note that \eqref{eq:trustr1} is a semidefinite program (SDP) and hence can be solved efficiently using, for example, the interior-point method \cite{Boyd_Convex04}.

Thus, for the MISO wiretap channel, the problem of finding the beamformer for the worst-case eavesdropper channel can be converted from a non-convex maximin problem into a quasiconvex problem that can be solved using the bisection method. Both the optimal covariance matrix and the corresponding worst-case channel mismatch are obtained. In the following section, the MISO wiretap channel with a cooperative jammer will be investigated.  
\section{Robust Cooperative Jamming} \label{sec:rcj}
We now consider the case when the Helper provides cooperative jamming to improve the secrecy rate.  According to the signal model in \eqref{eq:sigjam} and \eqref{eq:mismt}, the secrecy rate is
\begin{equation} \label{eq:srate}
R_s = \log_2 \left( 1+\frac{\mathbf{h}_b \mathbf{Q}_x \mathbf{h}_b^H}{\mathbf{g}_b \mathbf{Q}_z \mathbf{g}_b^H + \sigma^2} \right)
- \log_2 \left( 1+\frac{(\tilde{\mathbf{h}}_e+\mathbf{e}_h) \mathbf{Q}_x (\tilde{\mathbf{h}}_e+\mathbf{e}_h)^H}{ (\tilde{\mathbf{g}}_e+\mathbf{e}_g) \mathbf{Q}_z (\tilde{\mathbf{g}}_e+\mathbf{e}_g)^H + \sigma^2} \right).
\end{equation}
We will first consider the optimization problem under individual power constraints, \textit{i.e.}, $\mathbf{Q}_x \in \mathcal{Q}_x = \{\mathbf{Q}_x: \mathbf{Q}_x \succeq 0, {\textrm{tr}} (\mathbf{Q}_x) \le P_S\}$  and $\mathbf{Q}_z \in \mathcal{Q}_z = \{\mathbf{Q}_z: \mathbf{Q}_z \succeq 0, {\textrm{tr}} (\mathbf{Q}_z) \le P_J \}$, and then we investigate a more complicated case where a global power constraint $\textrm{tr}(\mathbf{Q}_x) + \textrm{tr}(\mathbf{Q}_z) \le P$ is imposed.

When the ECSI is perfectly known, maximization of $R_S$ over both $\mathbf{Q}_x$ and $\mathbf{Q}_z$ requires an iterative search \cite{Jorswieck_Secrecy10}. However, it has also been shown that performance close to the optimal solution for the MISO wiretap channel can be obtained by simply forcing the helper's signal to be orthogonal to Bob's channel. Therefore, to simplify the solution when the ECSI is imperfect, we use a zero-forcing (ZF) constraint on the jamming signal for the CJ problem, which is equivalent to requiring that $\mathbf{Q}_z \mathbf{g}_b^H = \mathbf{0}$. With the ZF constraint, the maximization of $R_S$ with respect to $\mathbf{Q}_z$ does not depend on $\mathbf{Q}_x$, although the optimal $\mathbf{Q}_x$ still depends on $\mathbf{Q}_z$. Thus, we will first optimize $\mathbf{Q}_z$ and then the optimal $\mathbf{Q}_x$ can be calculated.

\subsection{Individual Power Constraint}
For the case of perfect ECSI, the optimal $\mathbf{Q}_z$ under the ZF constraint is given by
\begin{subequations} \label{eq:nonrbf}
\begin{align}
\max_{\mathbf{Q}_z \in \mathcal{Q}_z}& \quad
\mathbf{g}_{e} \mathbf{Q}_z \mathbf{g}_{e}^H \\
\textrm{s.t.}& \quad \mathbf{g}_b \mathbf{Q}_z \mathbf{g}_b^H = 0.
\end{align}
\end{subequations}

\begin{lemma} \label{th:rank1}
The optimal covariance matrix $\mathbf{Q}_z \in \mathcal{Q}_z$ for problem \eqref{eq:nonrbf} is rank one.
\end{lemma}
\begin{IEEEproof}
See Appendix \ref{sec:rank1pf}.
\end{IEEEproof}

According to Lemma \ref{th:rank1}, the optimal ZF jamming signal for the perfect ECSI case is also single-stream beamforming, \textit{i.e.} $\mathbf{Q}_z=P_J \mathbf{w} \mathbf{w}^H$, where $\mathbf{w}$ is the unit-normalized one-dimensional beamformer for the Helper. We can rewrite problem
\eqref{eq:nonrbf} as
\begin{subequations} \label{rayl}
\begin{align}
\max_{\mathbf{w}}& \quad |\mathbf{g}_{e} \mathbf{w}|^2  \\
{\textrm{s.t.}}& \quad \mathbf{g}_b \mathbf{w} = 0 \\
           & \quad \mathbf{w}^H \mathbf{w} = 1.
\end{align}
\end{subequations}
The solution for problem \eqref{rayl} is referred to as the null-steering beamformer and is given by \cite{Friedlander_Performance89}
\begin{equation} \label{eq:wnull}
\mathbf{w}^* = \frac{(\mathbf{I}_{N_h} - \mathbf{P}_{gb}) \mathbf{g}_e^H}{||(\mathbf{I}_{N_h} - \mathbf{P}_{gb}) \mathbf{g}_e^H||}
\end{equation}
where $\mathbf{P}_{gb}=\mathbf{g}_b^H (\mathbf{g}_b \mathbf{g}_b^H)^{-1} \mathbf{g}_b$ is the orthogonal projection onto the subspace spanned by $\mathbf{g}_b^H$. The optimal information covariance matrix $\mathbf{Q}_x$, similar to the perfect ECSI case discussed in Section \ref{sec:rdt}, is rank one and the corresponding beamformer is the generalized eigenvector of the matrix pencil $(\sigma^2 \mathbf{I} + P_S \mathbf{h}_{b}^H \mathbf{h}_{b}, \sigma_z^2 \mathbf{I} + P_S \mathbf{h}_{e}^H \mathbf{h}_{e})$ with the largest generalized eigenvalue, where $\sigma_z^2 = \sigma^2+\mathbf{g}_{e} \mathbf{Q}_z \mathbf{g}_{e}^H$.

For the case of imperfect ECSI, we still solve for the jamming covariance first, and the optimization problem becomes
\begin{subequations} \label{eq:rjamor}
\begin{align}
\max_{\mathbf{Q}_z \in \mathcal{Q}_z} \min_{\mathbf{e}_g \in \mathcal{E}_g}& \quad
(\tilde{\mathbf{g}}_e+\mathbf{e}_g) \mathbf{Q}_z (\tilde{\mathbf{g}}_e+\mathbf{e}_g)^H \\
\textrm{s.t.}& \quad \mathbf{g}_b \mathbf{Q}_z \mathbf{g}_b^H = 0.
\end{align}
\end{subequations}
\begin{proposition} \label{th:maxqz}
Problem \eqref{eq:rjamor} is equivalent to the following problem
\begin{subequations} \label{eq:prop2}
\begin{align}
\max_{\mathbf{Q}_z, \mu, \mathbf{\Psi}}& \quad
\textrm{tr} [(\mathbf{Q}_z-\mathbf{\Psi}) \tilde{\mathbf{g}}_e^H \tilde{\mathbf{g}}_e]-\mu \epsilon_g^2 \\
{\textrm{s.t.}}& \quad
\left[
\begin{array}{cc}
\mathbf{\Psi} & \mathbf{Q}_z \\
\mathbf{Q}_z & \mu \mathbf{I}_{N_h} + \mathbf{Q}_z
\end{array}
\right ] \succeq 0 \\
					 & \quad   \textrm{tr}(\mathbf{Q}_z) \le P_J \\
					 & \quad   \mathbf{Q}_z \succeq 0, \mu \ge 0 \\
                     & \quad \mathbf{g}_b \mathbf{Q}_z \mathbf{g}_b^H = 0.
\end{align}
\end{subequations}
\end{proposition}
\begin{IEEEproof}
See Appendix \ref{sec:maxqzpf}.
\end{IEEEproof}

Problem \eqref{eq:prop2} is an SDP that consists of a linear objective function, together with a set of LMI constraints. Therefore, we can solve this problem efficiently and obtain the optimal solution $\mathbf{Q}_z^*$. Note that although $\mathbf{e}_g$ does not explicitly appear in \eqref{eq:prop2}, the optimal robust covariance $\mathbf{Q}_z^*$ is already based on the hidden worst-case $\mathbf{e}_g^*$ that can be expressed explicitly through the following problem
\begin{subequations} \label{eq:maxeg}
\begin{align}
\min_{\mathbf{e}_g}& \quad (\tilde{\mathbf{g}}_e+\mathbf{e}_g) \mathbf{Q}_z^* (\tilde{\mathbf{g}}_e+\mathbf{e}_g)^H \\
\textrm{s.t.}& \quad ||\mathbf{e}_g|| \le \epsilon_g.
\end{align}
\end{subequations}
This problem is similar to \eqref{eq:maxeh} with the difference that \eqref{eq:maxeg} is a convex problem and thus strong duality holds for \eqref{eq:maxeg} and its dual. The worst-case channel mismatch is provided through the following proposition.
\begin{proposition} \label{th:weg}
The worst channel mismatch for problem \eqref{eq:maxeg} is given by $\mathbf{e}_g^* = -\tilde{\mathbf{g}}_e \mathbf{Q}_z^* (\lambda \mathbf{I} + \mathbf{Q}_z^*)^{-1}$, where $\lambda$ is the solution of the following SDP problem
\begin{subequations}
\begin{align}
\max_{\lambda \ge 0, \gamma}& \quad \gamma \\
{\textrm{s.t.}}& \quad
\left[
\begin{array}{cc}
\lambda \mathbf{I} + \mathbf{Q}_z^* & \mathbf{Q}_z^* \tilde{\mathbf{g}}_e^H \\
\tilde{\mathbf{g}}_e \mathbf{Q}_z^* & \tilde{\mathbf{g}}_e \mathbf{Q}_z^* \tilde{\mathbf{g}}_e^H - \lambda \epsilon_g^2 - \gamma
\end{array}
\right ] \succeq 0.
\end{align}
\end{subequations}
\end{proposition}

\begin{IEEEproof}
See Appendix \ref{sec:wegpf}.
\end{IEEEproof}

With solutions for $\mathbf{Q}_z^*$ and $\mathbf{e}_g^*$, we can follow \eqref{eq:maximin1}-\eqref{eq:lfp} and formulate the optimization problem over $\mathbf{Q}_x$ as
\begin{equation} \label{eq:maximin2}
\max_{\mathbf{Q}_x \in \mathcal{Q}_x} \min_{\mathbf{e}_h \in \mathcal{E}_h}
\frac{\sigma^2+(\tilde{\mathbf{g}}_e+\mathbf{e}_g^*) \mathbf{Q}_z^* (\tilde{\mathbf{g}}_e+\mathbf{e}_g^*)^H+\mathbf{h}_b \mathbf{Q}_x \mathbf{h}_b^H}
{\sigma^2+ (\tilde{\mathbf{h}}_e+\mathbf{e}_h) \mathbf{Q}_x (\tilde{\mathbf{h}}_e+\mathbf{e}_h)^H},
\end{equation}
which can be solved with the same procedure as in Section \ref{sec:rdt}.

\subsection{Global Power Constraint} \label{sec:globalp}
The previous sections only consider the design of the information covariance $\mathbf{Q}_x$ and the jamming covariance $\mathbf{Q}_z$ under the assumption that their individual power constraints $P_S$ and $P_J$ are fixed. This section investigates the joint optimization over $\mathbf{Q}_x$, $\mathbf{Q}_z$ and the power allocation between Alice and the Helper, under the constraint that $\textrm{tr}(\mathbf{Q}_x) + \textrm{tr}(\mathbf{Q}_z) = p_1 + p_2 \le P$. As with the previous case, we will assume a zero-forcing constraint for the helper's jamming signal at Bob. We will first optimize the power allocation for a pair of given $\mathbf{Q}_x$ and $\mathbf{Q}_z$, and then we will provide an iterative algorithm that updates both the beamformers and the power allocation.

First, for given $\mathbf{Q}_x$ and $\mathbf{Q}_z$, let $\mathbf{Q}_x = p_1 \bar{\mathbf{Q}}_x$ and $\mathbf{Q}_z = p_2 \bar{\mathbf{Q}}_z$ where $\bar{\mathbf{Q}}_x$ and $\bar{\mathbf{Q}}_z$ are normalized such that $\textrm{tr}(\bar{\mathbf{Q}}_x)=1$ and $\textrm{tr}(\bar{\mathbf{Q}}_z)=1$. Hence the maximization of the secrecy rate  \eqref{eq:srate} with respect to $p_1$ and $p_2$ is equivalent to
\begin{subequations} \label{eq:pagp}
\begin{align}
\max_{p_1,p_2 \ge 0}& \quad \frac{p_1 p_2 c_1 c_3 + p_1 c_1 \sigma^2 + p_2 c_3 \sigma^2 + \sigma^4}{p_1 c_2 + p_2 c_3 + \sigma^2} \label{eq:qfun} \\
\textrm{s.t.}& \quad p_1 + p_2 \le P
\end{align}
\end{subequations}
where
\begin{align}
c_1 =& \mathbf{h}_b \bar{\mathbf{Q}}_x \mathbf{h}_b^H \\
c_2 =& (\tilde{\mathbf{h}}_e+\mathbf{e}_h) \bar{\mathbf{Q}}_x (\tilde{\mathbf{h}}_e+\mathbf{e}_h)^H \\
c_3 =&  (\tilde{\mathbf{g}}_e+\mathbf{e}_g) \bar{\mathbf{Q}}_z (\tilde{\mathbf{g}}_e+\mathbf{e}_g)^H.
\end{align}
Since \eqref{eq:qfun} is a quadratic fractional function, the optimization problem is hard to solve directly. However, we can use the \textit{single condensation method} to solve this non-convex problem via a series of geometric programming (GP) steps \cite{Chiang_Power07}. GP is a class of non-linear optimization problems that can be readily turned into convex optimization problems, and hence a global optimum can be efficiently computed \cite{Boyd_tutorial07}. Before applying the single condensation method, we give the following lemma \cite{Chiang_Power07}:
\begin{lemma} \label{th:approx}
Given a posynomial
\begin{equation}
f(\mathbf{x})=\sum_{i=1}^{m} u_i(\mathbf{x})=\sum_{i=1}^{m} c_i x_1^{\beta_{1i}}x_2^{\beta_{2i}} \cdots x_{n}^{\beta_{ni}},
\end{equation}
where the exponents $\beta_{ji}$ are arbitrary real numbers and $c_i$ are positive constants, the following inequality holds:
\begin{equation}
f(\mathbf{x}) \ge  \tilde{f}(\mathbf{x})
= \prod_{i=1}^{m} \left(\frac{u_i(\mathbf{x})}{\alpha_i}\right)^{\alpha_i},
\end{equation}
where $\alpha_i>0$ and $\sum_{i=1}^{m} \alpha_i = 1$. The inequality becomes an equality when $\alpha_i = \frac{u_i(\mathbf{x_0})}{f(\mathbf{x_0})}$,
in which case the monomial $\tilde{f}(\mathbf{x_0})$ is the best local approximation of the posynomial $f(\mathbf{x_0})$ near $\mathbf{x_0}$.
\end{lemma}
\begin{IEEEproof}
We can rewrite $f(\mathbf{x})$ as
\begin{align}
f(\mathbf{x}) &= \sum_{i=1}^{m} \alpha_i \left( \frac{u_i(\mathbf{x})}{\alpha_i} \right) \\
 							& \ge \prod_{i=1}^{m} \left(\frac{u_i(\mathbf{x})}{\alpha_i}\right)^{\alpha_i} \label{eq:condg}
\end{align}
where \eqref{eq:condg} holds according to the arithmetic-geometric mean inequality. When $\alpha_i = \frac{u_i(\mathbf{x_0})}{f(\mathbf{x_0})}$, we have that $\alpha_i$ $(i = 1,\cdots,m)$ are all positive coefficients with $\sum_{i=1}^{m} \alpha_i = 1$, and the proof of equality is straightforward by inserting $\alpha_i$ back into $\tilde{f}(\mathbf{x_0})$.
\end{IEEEproof}

Using Lemma \ref{th:approx}, let
\begin{equation} \label{eq:f1}
f(p_1,p_2) = p_1 p_2 c_1 c_3 + p_1 c_1 \sigma^2 + p_2 c_3 \sigma^2 + \sigma^4,
\end{equation}
and rewrite the numerator of \eqref{eq:qfun} as
\begin{align}
f(p_1,p_2) =& \tilde{f}(p_1,p_2) \notag \\
											=&\left( \frac{p_1 p_2 c_1 c_3}{\alpha_1} \right)^{\alpha_1} \left( \frac{p_1 c_1 \sigma^2}{\alpha_2} \right)^{\alpha_2}  \left( \frac{p_2 c_3 \sigma^2}{\alpha_3} \right)^{\alpha_3} \left( \frac{\sigma^4}{\alpha_4} \right)^{\alpha_4} \label{eq:condf}
\end{align}
where
\begin{align}
\alpha_1 &= \frac{p_1 p_2 c_1 c_3}{f(p_1,p_2)} \label{eq:apco} \\
\alpha_2 &= \frac{p_1 c_1 \sigma^2}{f(p_1,p_2)} \\
\alpha_3 &= \frac{p_2 c_3 \sigma^2}{f(p_1,p_2)}\\
\alpha_4 &= \frac{\sigma^4}{f(p_1,p_2)}, \label{eq:btco}
\end{align}
so the optimization problem of \eqref{eq:pagp} becomes
\begin{subequations} \label{eq:gp}
\begin{align}
\min_{p_1,p_2 \ge 0}& \quad \frac{p_1 c_2 + p_2 c_3 + \sigma^2}
{\tilde{f}_2(p_1,p_2)} \label{eq:gp1} \\
\textrm{s.t.}& \quad p_1 + p_2 \le P. \label{eq:gp2}
\end{align}
\end{subequations}
The optimization problem stated above can be readily converted into the standard form for geometric programming problems, and \eqref{eq:gp1}-\eqref{eq:gp2} are posynomials of the GP problem. Therefore, the global optimal solution can be efficiently obtained for problem \eqref{eq:gp}. Next, we can use the single condensation method to solve problem \eqref{eq:pagp}, and the main steps are outlined in Algorithm \ref{al:scm}.

\begin{pseudocode}[framebox]{Single condensation method for global power allocation}{} \label{al:scm}
\INIT \TX{power allocation } p_1^{(0)} \TX{ and } p_2^{(0)}. \\
\ITER \\
\DO
\BEGIN
\TX{For iteration } k, \TX{evaluate posynomial } f(p_1^{(k-1)},p_2^{(k-1)}) \TX{ according to \eqref{eq:f1}}. \\
\TX{Compute } \alpha_i^{(k)} \TX{according to \eqref{eq:apco}-\eqref{eq:btco} using } p_1^{(k-1)}, p_2^{(k-1)}.\\
\TX{Condense posynomial } f \TX{ into monomial } \tilde{f} \TX{ according to \eqref{eq:condf}}.\\
\TX{Solve problem \eqref{eq:gp} with a single GP.} \\
\TX{Apply the resulting } p_1^{(k)} \TX{ and } p_2^{(k)} \TX{ into the first step and loop until convergence.}
\END
\end{pseudocode}\\
This successive optimization method is based on GP problems that can be solved using interior-point
methods with polynomial-time complexity, and it has been proven in \cite{Chiang_Power07} that the solution obtained using successive approximations for the single condensation method will efficiently converge to a point satisfying the KKT conditions of the original problem.

Now we can conduct the joint optimization that considers both the information/jamming covariances and the power allocation between them. The main steps are outlined as follows:
\begin{enumerate}
\item Initilize $p_1=p_2=\frac{P}{2}$.
\item Let $P_S=p_1$, $P_J=p_2$ and solve problem \eqref{eq:rjamor}, \eqref{eq:maximin2} to obtain $\mathbf{Q}_x^*$, $\mathbf{Q}_z^*$, $\mathbf{e}_h^*$ and $\mathbf{e}_g^*$.
\item Let $\bar{\mathbf{Q}}_x= \frac{\mathbf{Q}_x}{\textrm{tr}(\mathbf{Q}_x)}, \bar{\mathbf{Q}}_z= \frac{\mathbf{Q}_z}{ \textrm{tr}(\mathbf{Q}_z)}$ and solve problem \eqref{eq:pagp} with Algorithm \ref{al:scm}.
\item Apply the resulting $p_1$ and $p_2$ to step 2 and loop until convergence.
\end{enumerate}
Since in each iteration the objective functions
are maximized and the secrecy rate is increased, and since there is an upper bound for the secrecy rate (the case without an eavesdropper where the jamming power is zero), the procedure will converge to an optimum. Our extensive numerical experiments, some results of which are shown in Section \ref{sec:nr}, further illustrate that the global optimum is obtained through this procedure. Note that the non-robust counterpart of this procedure (where the CSI is assumed to be perfect) is similar but with the difference that instead of obtaining the covariance matrices from the robust maximin problems \eqref{eq:rjamor} and \eqref{eq:maximin2} in step 2, $\mathbf{Q}_x^*$ and $\mathbf{Q}_z^*$ are the solutions of the non-robust problems in \eqref{eq:nonrbf}-\eqref{eq:wnull}.

\section{Robust Transmit Design with QoS Constraint} \label{sec:qos}
In this section, we consider a slightly different problem in which there is a fixed constraint on the signal to interference-plus-noise ratio (SINR) of the legitimate link, so that the problem reduces to minimizing the SINR at the eavesdropper for the worst channel mismatch, under a given power constraint. As we will see, the addition of the SINR constraint actually simplifies the robust solution.

\subsection{Robust Direct Transmission} \label{sec:qosrdt}
When the Helper is silent, the optimization problem for the perfect-ECSI case is
\begin{subequations} \label{eq:qosdt}
\begin{align}
\min_{\mathbf{Q}_x \in \mathcal{Q}_x}& \quad \mathbf{h}_e \mathbf{Q}_x \mathbf{h}_e^H \\
\textrm{s.t.}& \quad \frac{\mathbf{h}_b \mathbf{Q}_x \mathbf{h}_b^H}{\sigma^2} \ge \gamma_t,
\end{align}
\end{subequations}
where $\gamma_t$ is the desired target SINR at Bob. When the channel mismatch is considered, the optimization problem is formulated as
\begin{subequations} \label{eq:qosrdt}
\begin{align}
\min_{\mathbf{Q}_x \in \mathcal{Q}_x} \max_{\mathbf{e}_h \in \mathcal{E}_h}& \quad (\tilde{\mathbf{h}}_e+\mathbf{e}_h) \mathbf{Q}_x (\tilde{\mathbf{h}}_e+\mathbf{e}_h)^H \\
\textrm{s.t.}& \quad \frac{\mathbf{h}_b \mathbf{Q}_x \mathbf{h}_b^H}{\sigma^2} \ge \gamma_t.
\end{align}
\end{subequations}
Comparing problem \eqref{eq:qosrdt} with \eqref{eq:maximin1}, we see that \eqref{eq:qosrdt} simplifies the fractional expression in \eqref{eq:maximin1} by introducing an extra affine inequality constraint. Therefore, the procedure for solving \eqref{eq:maximin1} can be applied to solve \eqref{eq:qosrdt}, and we have the similar result that \eqref{eq:qosrdt} is equivalent to the following optimization problem:
\begin{subequations} \label{eq:qosrdtprop}
\begin{align}
\min_{\mathbf{Q}_x, \mu, \mathbf{\Psi}}& \quad
\mu \epsilon_h^2 + \textrm{tr} [(\mathbf{Q}_x+\mathbf{\Psi}) \tilde{\mathbf{h}}_e^H \tilde{\mathbf{h}}_e] \\
{\textrm{s.t.}}& \quad
\left[
\begin{array}{cc}
\mathbf{\Psi} & \mathbf{Q}_x \\
\mathbf{Q}_x & \mu \mathbf{I}_{N_a} -\mathbf{Q}_x
\end{array}
\right ] \succeq 0 \\
					 & \quad   \textrm{tr}(\mathbf{Q}_x) \le P \\
					 & \quad   \mathbf{Q}_x \succeq 0, \mu \ge 0 \\
                     & \quad   \textrm{tr}(\mathbf{Q}_x \mathbf{h}_b^H \mathbf{h}_b) \ge \sigma^2 \gamma_t.
\end{align}
\end{subequations}
Unlike the quasiconvex problem in Proposition \ref{th:maxqx} that requires a bisection method, \eqref{eq:qosrdtprop} is an SDP with a linear objective function, and thus can be solved efficiently.

Note that by exploiting a similar form in \eqref{eq:eqv}, problem \eqref{eq:qosrdt} is also equivalent to
\begin{subequations} \label{eq:relqx}
\begin{align}
\min_{\mathbf{Q}_x \in \mathcal{Q}_x, \mu \ge ||\mathbf{Q}_x||_2}& \quad \mu \epsilon_h^2 + \tilde{\mathbf{h}}_e \mathbf{Q}_x \tilde{\mathbf{h}}_e^H + \tilde{\mathbf{h}}_e \mathbf{Q}_x
(\mu \mathbf{I}_{N_a} - \mathbf{Q}_x)^\dag \mathbf{Q}_x \tilde{\mathbf{h}}_e^H \\
\textrm{s.t.}& \quad \frac{\mathbf{h}_b \mathbf{Q}_x \mathbf{h}_b^H}{\sigma^2} \ge \gamma_t.
\end{align}
\end{subequations}
Denoting $\tilde{\mathbf{Q}}_x = \frac{\mathbf{Q}_x}{\textrm{tr}(\mathbf{Q}_x)}$, and based on the observation that $\tilde{\mathbf{h}}_e \mathbf{Q}_x \tilde{\mathbf{h}}_e^H \ge 0$, $\tilde{\mathbf{h}}_e \mathbf{Q}_x
(\mu \mathbf{I}_{N_a} - \mathbf{Q}_x)^\dag \mathbf{Q}_x \tilde{\mathbf{h}}_e^H \ge 0$, a relaxed solution for $\mathbf{Q}_x$ of \eqref{eq:relqx} can be obtained by steering the eigenvectors of $\mathbf{Q}_x$ to maximize $\mathbf{h}_b \tilde{\mathbf{Q}}_x \mathbf{h}_b^H$ subject to a zero-forcing constraint $\tilde{\mathbf{Q}}_x \mathbf{h}_e^H = \mathbf{0}$, while allocating the eigenvalues of $\mathbf{Q}_x$ to satisfy the SINR constraint. Thus the relaxed optimization becomes
\begin{subequations} \label{eq:relqx1}
\begin{align}
\max_{\tilde{\mathbf{Q}}_x}& \quad \mathbf{h}_b \tilde{\mathbf{Q}}_x \mathbf{h}_b^H \\
\textrm{s.t.}& \quad \tilde{\mathbf{h}}_e \tilde{\mathbf{Q}}_x \tilde{\mathbf{h}_e}^H = 0,
\end{align}
\end{subequations}
and $\mathbf{Q}_x^* = \frac{\sigma^2 \gamma_t}{\mathbf{h}_b \tilde{\mathbf{Q}}_x^* \mathbf{h}_b^H} \tilde{\mathbf{Q}}_x^* $. Note that \eqref{eq:relqx1} is also a relaxed optimization problem for its non-robust counterpart in \eqref{eq:qosdt}, because when there is enough transmit power to meet the SINR demand, Alice will choose a beamformer that will null the signal at Eve while aligning it as close to $\mathbf{h}_b^H$ as possible. Therefore, in this case, the non-robust and robust covariance matrices, and thus the corresponding worst-case SINRs, are expected to be very close to each other. This will be illustrated by the simulation results of Section \ref{sec:nr}.

\subsection{Robust Cooperative Jamming}
Unlike the previous section, in the cooperative jamming case with a QoS constraint, we will not need to introduce a ZF constraint on the jamming signal from the helper to simplify the problem. Note also that we only consider the global power constraint scenario, since the generalization of the proposed solution to the case of individual power constraints is straightforward.

For the joint optimization of $\mathbf{Q}_x$ and $\mathbf{Q}_z$ under the QoS constraint, the optimization problem is given by
\begin{subequations} \label{eq:qosrcj}
\begin{align}
\min_{\mathbf{Q}_x, \mathbf{Q}_z} \max_{\mathbf{e}_h, \mathbf{e}_g}& \quad
\frac{(\tilde{\mathbf{h}}_e+\mathbf{e}_h) \mathbf{Q}_x (\tilde{\mathbf{h}}_e+\mathbf{e}_h)^H}{ (\tilde{\mathbf{g}}_e+\mathbf{e}_g) \mathbf{Q}_z (\tilde{\mathbf{g}}_e+\mathbf{e}_g)^H + \sigma^2} \\
\textrm{s.t.}& \quad \textrm{tr}(\mathbf{Q}_x) + \textrm{tr}(\mathbf{Q}_z) \le P \\
             & \quad |\mathbf{e}_h| \le \epsilon_h, |\mathbf{e}_g| \le \epsilon_g \\
             & \quad \frac{\mathbf{h}_b \mathbf{Q}_x \mathbf{h}_b^H}{\mathbf{g}_b \mathbf{Q}_z \mathbf{g}_b^H + \sigma^2} \ge \gamma_t.
\end{align}
\end{subequations}
Unlike the global power allocation problem discussed in Section \ref{sec:globalp}, the QoS constraint in problem \eqref{eq:qosrcj} simplifies the fractional quadratic expression in \eqref{eq:pagp} into a linear fractional form with respect to $\mathbf{Q}_x$ and $\mathbf{Q}_z$. Thus we can directly obtain the solution via the following proposition.
\begin{proposition} \label{th:qosrcjp}
Problem \eqref{eq:qosrcj} is equivalent to the following problem
\begin{subequations} \label{eq:qosrcjp}
\begin{align}
\min_{\mathbf{Q}_x, \mathbf{Q}_z, \mathbf{\Psi}, \mathbf{\Phi}, \mu, \nu}& \quad
\frac{\mu \epsilon_h^2 + \textrm{tr} [(\mathbf{Q}_x+\mathbf{\Psi}) \tilde{\mathbf{h}}_e^H \tilde{\mathbf{h}}_e]} {\sigma^2 - \nu \epsilon_g^2 + \textrm{tr} [(\mathbf{Q}_z-\mathbf{\Phi}) \tilde{\mathbf{g}}_e^H \tilde{\mathbf{g}}_e]} \\
{\textrm{s.t.}}& \quad
\left[
\begin{array}{cc}
\mathbf{\Psi} & \mathbf{Q}_x \\
\mathbf{Q}_x & \mu \mathbf{I}_{N_a} - \mathbf{Q}_x
\end{array}
\right ] \succeq 0 \\
& \quad
\left[
\begin{array}{cc}
\mathbf{\Phi} & \mathbf{Q}_z \\
\mathbf{Q}_z & \nu \mathbf{I}_{N_h} + \mathbf{Q}_z
\end{array}
\right ] \succeq 0 \\
                     & \quad   \textrm{tr}(\mathbf{Q}_x)+\textrm{tr}(\mathbf{Q}_z) \le P \\
					 & \quad   \mathbf{Q}_x \succeq 0,\mathbf{Q}_z \succeq 0, \mu \ge 0,\nu \ge 0 \\
                     & \quad \textrm{tr}( \mathbf{Q}_x \mathbf{h}_b^H \mathbf{h}_b)
                     \ge \gamma_t(\textrm{tr}( \mathbf{Q}_z \mathbf{g}_b^H \mathbf{g}_b) + \sigma^2).
\end{align}
\end{subequations}
\end{proposition}
\begin{IEEEproof}
See Appendix \ref{sec:appqosrcjp}.
\end{IEEEproof}

Problem \eqref{eq:qosrcjp} is also an SDP with a quasiconvex objective function, a set of LMIs and affine inequalities. Thus it can be solved via the bisection method discussed in Section \ref{sec:rdt}. Note that for some cases when the SINR requirement at Bob can not be met for a given power constraint, the optimization problem will not be feasible and the transmission is assumed to be in outage. For the non-robust counterpart of the above QoS-constrained cooperative jamming problem, Alice will still use the solution of problem \eqref{eq:qosdt}, and the Helper will use the remainder of the power for ZF jamming (same as problem \eqref{eq:nonrbf}), since the Helper has no information about the channel mismatch. 
\section{Numerical Results} \label{sec:nr}
In this section, we present some numerical examples on the secrecy rate performance of the robust transmission schemes studied in the paper. For all examples, we assume Alice and the Helper both have four antennas, \textit{i.e.} $N_a=N_h=4$,  while Bob and Eve each has one. The channel matrices are assumed to be composed of independent, zero-mean Gaussian random variables with unit variance. All
results are calculated based on an average of 1000 independent
trials. The background noise power is assumed to be the same at Bob and Eve, $\sigma_b^2=\sigma_e^2=1$, and the transmit power $P$ is defined in dB relative to the noise power.

We will examine the performance of the robust direct transmission (DT) scheme and the robust cooperative jamming (CJ) scheme under various power constraints, channel error bounds and QoS constraints. For purposes of comparison, we also examine the non-robust generalized eigenvector schemes (which will be referred to as GEV DT) discussed in Section \ref{sec:rdt} , and the non-robust DT and CJ schemes discussed in Sections \ref{sec:rdt}-\ref{sec:qos}.

\begin{figure}[ht]
\begin{center}
\includegraphics[width=0.68\textwidth]{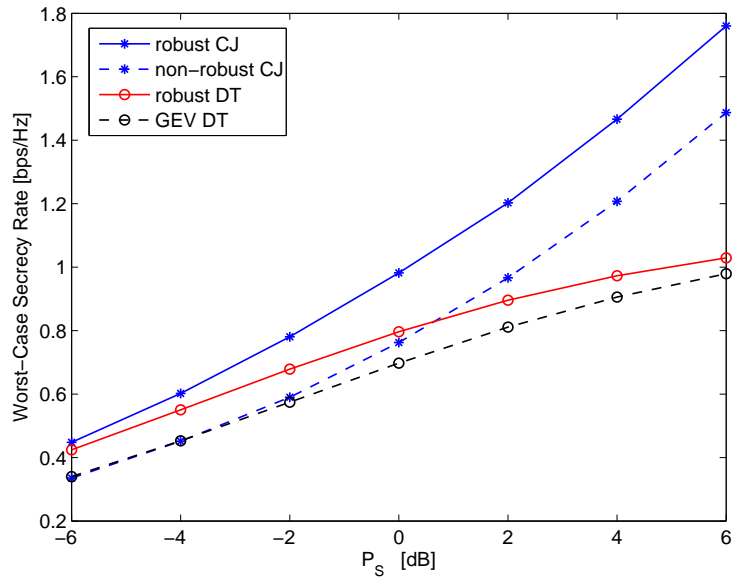}
\caption{\label{fig:sridp} Worst-case secrecy rate vs. transmit power, with individual power constraint, $\epsilon_h^2=\epsilon_g^2=1.5$.}
\end{center}
\end{figure}

Fig.~\ref{fig:sridp} shows the worst-case secrecy rate as a function of transmit power under an individual power constraint, assuming $P_S=P_J$ and the channel mismatch is $\epsilon_h^2=\epsilon_g^2=1.5$. In general, when there exists channel mismatch, we see that the robust design for DT and CJ produces better performance in terms of the worst-case secrecy rate compared to their non-robust counterparts. Note that the non-robust CJ scheme, although possessing twice the available power as that for DT, performs even worse than the robust DT scheme for low transmit powers. This is due to the fact that the performance of the non-robust CJ scheme is degraded not only by the channel error between Alice and Eve, but also by that between the Helper and Eve.

\begin{figure}[ht]
\begin{center}
\includegraphics[width=0.68\textwidth]{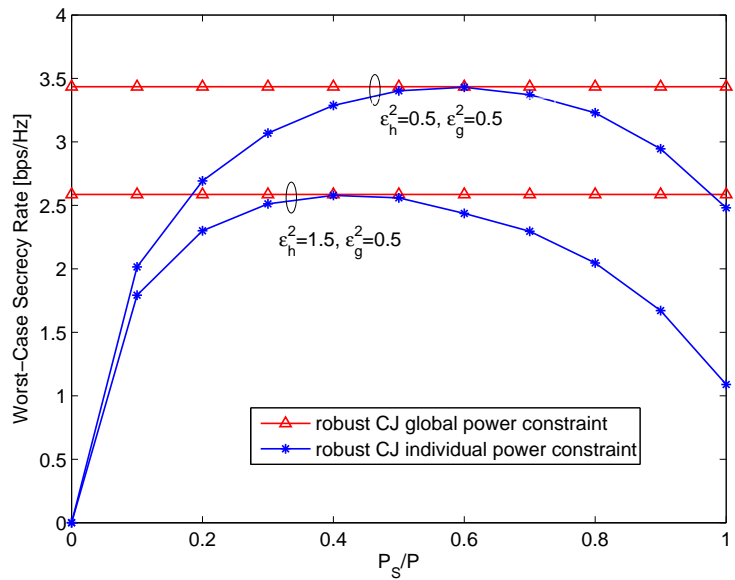}
\caption{\label{fig:gp} Worst-case secrecy rate vs. transmit power fraction, $P=10$dB.}
\end{center}
\end{figure}

In Fig.~\ref{fig:gp}, we compare the performance of the robust CJ scheme under both global and individual power constraints. In this case, we assume the global power limit $P$ is $10$dB, $P_S$ and $P_J$ are the individual power constraints for Alice and the Helper respectively, and $P_S+P_J=P$. The benefit of having the flexibility associated with a global power constraint over fixed individual power
constraint is clearly evident. Also it can be seen that the proposed joint optimization procedure for the global power constraint achieves the optimal worst-case secrecy rate. When $\epsilon_h^2$ increases, a larger fraction of the transmit power must be devoted to jamming in order to reach the higher secrecy rate.

\begin{figure}[ht]
\begin{center}
\includegraphics[width=0.68\textwidth]{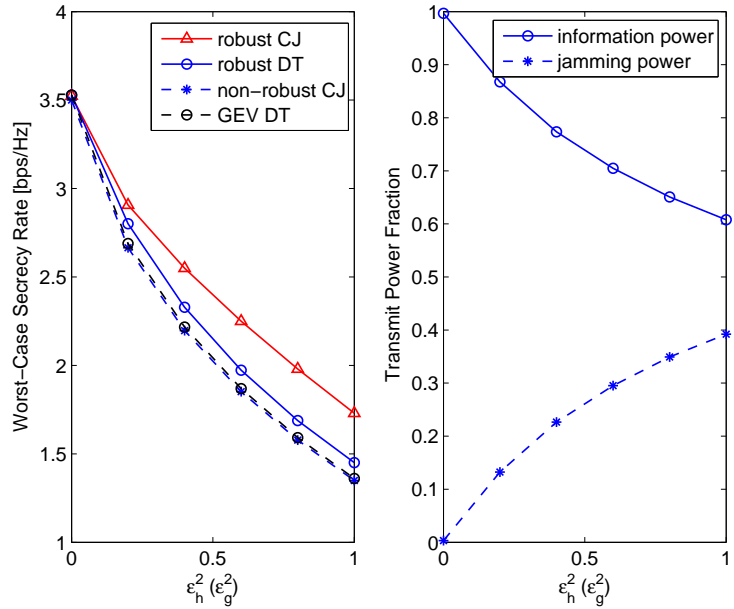}
\caption{\label{fig:srerr} Worst-case secrecy rate and transmit power fraction vs. channel mismatch, $P=5$dB.}
\end{center}
\end{figure}

The impact of the channel mismatch on the secrecy rate of the different schemes is presented in Fig.~\ref{fig:srerr}. The transmit power fraction for the robust CJ scheme is also plotted, and a global power constraint is used in this case. We assume $P$ is $5$dB, and the channel mismatch $\epsilon_h^2=\epsilon_g^2$. It can be observed that when the channel mismatch is zero, a jamming signal is not necessary, and all schemes achieve the same secrecy rate. However, when $\epsilon_h^2$ and $\epsilon_g^2$ increase, the robustness of the CJ scheme is more obvious, and the jamming fraction of the total transmit power also increases. Also it can be seen that the worst-case secrecy rate is much lower for the non-robust CJ scheme due to the fact that it is impacted by the channel mismatch from both the link between Alice and Eve, and the link between the Helper and Eve.

\begin{figure}[ht]
\begin{center}
\includegraphics[width=0.68\textwidth]{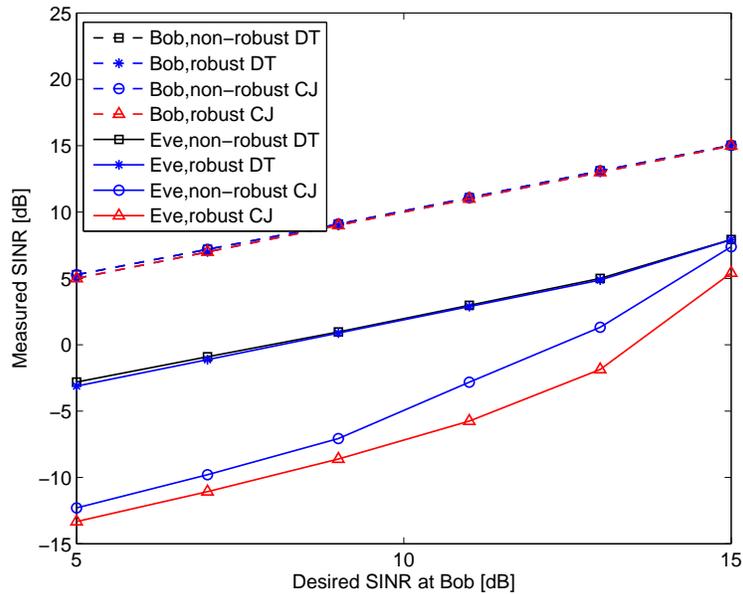}
\caption{\label{fig:qosbob} Worst-case secrecy rate vs. SINR constraint at Bob, $\epsilon_h^2=\epsilon_g^2=0.5$, $P=10$dB.}
\end{center}
\end{figure}

Next, we consider examples for the case where a desired SINR constraint is imposed at Bob. $P$ is assumed to be $10$dB in these examples. In Fig.~\ref{fig:qosbob}, the channel mismatch is given by $\epsilon_h^2=\epsilon_g^2=0.5$, and we plot the measured SINR at Bob and Eve with an increasing SINR constraint at Bob. Since the SINR constraint is met in all cases, the curves for Bob all coincide. We see that in this QoS constraint scenario, the robust CJ scheme still shows a significant performance improvement by suppressing the SINR at Eve. An interesting observation is that the robust DT scheme for minimizing the worst-case SINR at Eve shows almost the same performance as the non-robust DT method, which is consistent with the analytical prediction discussed in Section \ref{sec:qosrdt}. Therefore, the worst-case optimization for DT is unnecessary in this case. For the CJ schemes, when the SINR requirement at Bob is higher, more power will be allocated to Alice and less power will be used for jamming. Therefore, the performance of the CJ schemes approaches that of DT schemes.

\begin{figure}[ht]
\begin{center}
\includegraphics[width=0.68\textwidth]{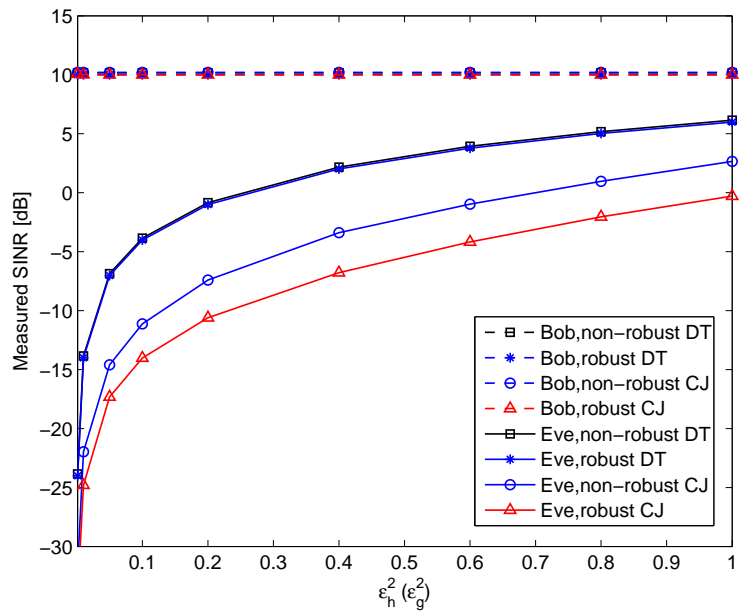}
\caption{\label{fig:qoserr} Worst-Case secrecy rate vs. channel mismatch, with QoS constraint $\gamma_t=10$dB, $P=10$dB.}
\end{center}
\end{figure}

In Fig.~\ref{fig:qoserr}, the impact of increasing $\epsilon_h^2$ and $\epsilon_g^2$ is depicted. In this example, the desired SINR constraint at Bob is set to be $10$dB. We see that when no channel mismatch exists, all schemes give zero SINR at Eve while maintaining the QoS constraint at Bob, and in such cases the use of artificial noise (CJ) is unnecessary. Also we can observe that although the SINR at Eve increases for all of the methods as the channel mismatch becomes larger, the benefits of using robust cooperative jamming over the other algorithms is obvious. 
\section{Conclusions} \label{sec:con}
In this paper, we studied robust transmit designs for MISO wiretap channels with imperfect ECSI. Robust transmit covariance matrices were obtained for both direct transmission and cooperative jamming scenarios, based on worst-case secrecy rate maximization. For the case of individual power constraints, we transformed the non-convex optimization problem into a quasiconvex problem. For the global power constraint case, we proposed an algorithm for joint optimization of the transmit covariance matrices and power allocation. In addition, we also obtained robust transmit covariance matrices for the scenario where a QoS constraint is imposed at the legitimate receiver. The benefits of the robust designs were illustrated  through numerical results. We conclude that although cooperative jamming is not helpful when perfect ECSI is available under a global power constraint, the worst-case secrecy rate can be increased and the SINR at Eve can be lowered by using jamming support from a helper when the ECSI is imperfect, provided that robust beamforming is employed.

\appendices
\section{Proof of Proposition 2} \label{sec:wehpf}
Problem \eqref{eq:maxeh} can be rewritten as
\begin{subequations} \label{eq:wstdel2}
\begin{align}
\min_{\mathbf{e}_h}& \quad - \mathbf{e}_h \mathbf{Q}_x^* \mathbf{e}_h^H - 2 \textrm{Re} (\tilde{\mathbf{h}}_e \mathbf{Q}_x^* \mathbf{e}_h^H) - \tilde{\mathbf{h}}_e \mathbf{Q}_x^* \tilde{\mathbf{h}}_e^H \\
\textrm{s.t.}& \quad \mathbf{e}_h \mathbf{e}_h^H \le \epsilon_h^2.
\end{align}
\end{subequations}
This is a non-convex problem since its Hessian is negative semidefinite, \textit{i.e.} $-\mathbf{Q}_x^* \preceq 0$. The Lagrangian of this problem is
\begin{align}
L(\mathbf{e}_h,\lambda) =& - \mathbf{e}_h \mathbf{Q}_x^* \mathbf{e}_h^H - 2 \textrm{Re} (\tilde{\mathbf{h}}_e \mathbf{Q}_x^* \mathbf{e}_h^H) - \tilde{\mathbf{h}}_e \mathbf{Q}_x^* \tilde{\mathbf{h}}_e^H + \lambda (\mathbf{e}_h \mathbf{e}_h^H - \epsilon_h^2) \notag \\
=& \mathbf{e}_h (\lambda \mathbf{I} - \mathbf{Q}_x^*) \mathbf{e}_h^H + 2 \textrm{Re} (-\tilde{\mathbf{h}}_e \mathbf{Q}_x^* \mathbf{e}_h^H) - \tilde{\mathbf{h}}_e \mathbf{Q}_x^* \tilde{\mathbf{h}}_e^H  - \lambda \epsilon_h^2
\end{align}
where $\lambda \ge 0$ and the dual function is given by
\begin{align}
&g(\lambda) = \inf_{\mathbf{e}_h} L(\mathbf{e}_h,\lambda) \notag \\
&= \left \{
\begin{array}{ll}
- \tilde{\mathbf{h}}_e \mathbf{Q}_x^* \tilde{\mathbf{h}}_e^H - \lambda \epsilon_h^2 -\tilde{\mathbf{h}}_e \mathbf{Q}_x^* (\lambda \mathbf{I} - \mathbf{Q}_x^*)^\dag \mathbf{Q}_x^*  \tilde{\mathbf{h}}_e^H  \quad &\lambda \mathbf{I} - \mathbf{Q}_x^* \succeq 0,   \mathbf{Q}_x^* \tilde{\mathbf{h}}_e^H \in \mathcal{R}(\lambda \mathbf{I} - \mathbf{Q}_x^*) \\
-\infty, & \textrm{otherwise}
\end{array} \right.
\end{align}
where the unconstrained minimization of $L(\mathbf{e}_h,\lambda)$ with respect to $\mathbf{e}_h$ is achieved when $\mathbf{e}_h = \tilde{\mathbf{h}}_e \mathbf{Q}_x^* (\lambda \mathbf{I} - \mathbf{Q}_x^*)^\dag$. The dual problem is thus
\begin{subequations}
\begin{align}
\max_{\lambda}& \quad - \tilde{\mathbf{h}}_e \mathbf{Q}_x^* \tilde{\mathbf{h}}_e^H - \lambda \epsilon_h^2 -\tilde{\mathbf{h}}_e \mathbf{Q}_x^* (\lambda \mathbf{I} - \mathbf{Q}_x^*)^\dag \mathbf{Q}_x^*  \tilde{\mathbf{h}}_e^H  \\
\textrm{s.t.}& \quad \lambda \mathbf{I} - \mathbf{Q}_x^* \succeq 0, \   \mathbf{Q}_x^* \tilde{\mathbf{h}}_e^H \in \mathcal{R}(\lambda \mathbf{I} - \mathbf{Q}_x^*).
\end{align}
\end{subequations}
Using a Schur complement, the dual problem becomes the following SDP
\begin{subequations} \label{eq:trustr}
\begin{align}
\max_{\lambda \ge 0, \gamma}& \quad \gamma \\
{\textrm{s.t.}}& \quad
\left[
\begin{array}{cc}
\lambda \mathbf{I} - \mathbf{Q}_x^* & \mathbf{Q}_x^* \tilde{\mathbf{h}}_e^H \\
\tilde{\mathbf{h}}_e \mathbf{Q}_x^* & - \tilde{\mathbf{h}}_e \mathbf{Q}_x^* \tilde{\mathbf{h}}_e^H - \lambda \epsilon_h^2 - \gamma
\end{array}
\right ] \succeq 0.
\end{align}
\end{subequations}
Note that \eqref{eq:wstdel2} is usually called a \textit{trust region subproblem} (TRS), and it has been proven that strong duality holds for TRS although the objective function is non-convex \cite{Stern_Indefinite95}. Thus the optimal value of \eqref{eq:wstdel2} and \eqref{eq:trustr} are the same.

\section{Proof of Lemma 1} \label{sec:rank1pf}
Since $\mathbf{g}_b \mathbf{Q}_z \mathbf{g}_b^H$ is nonnegative, we rewrite the equality constraint as $\mathbf{g}_b \mathbf{Q}_z \mathbf{g}_b^H \le \varepsilon$, where $\varepsilon>0$ is a arbitrarily small number. Denoting $f(\mathbf{Q}_z) = \frac{1}{\mathbf{g}_{e} \mathbf{Q}_z \mathbf{g}_{e}^H}$, the Lagrangian of problem \eqref{eq:nonrbf} is
\begin{equation}
L(\mathbf{Q}_z,\lambda,\mathbf{\Theta},\nu) = f(\mathbf{Q}_z)
 + \lambda (\textrm{tr}(\mathbf{Q}_z)- P_J) - \textrm{tr}(\mathbf{\Theta} \mathbf{Q}_z) + \nu \mathbf{g}_b \mathbf{Q}_z \mathbf{g}_b^H
\end{equation}
where $\lambda \ge 0$, $\nu \ge 0$, and $\mathbf{\Theta} \succeq 0$ is the Lagrange multiplier associated with the inequality constraint $\mathbf{Q}_z \ge 0$. Note that the objective function in \eqref{eq:nonrbf} is linear, and there exist strictly feasible points that satisfy both the inequality and equality constraints (\textit{e.g.} $\mathbf{Q}_z = \frac{P_J}{N_h} \mathbf{w} \mathbf{w}^H$, where $\mathbf{w}$ is a normalized vector orthogonal to $\mathbf{g}_b^H$). Therefore, according to Slater's theorem, the primal and dual optimal points of \eqref{eq:nonrbf} satisfy the Karush-Kuhn-Tucker (KKT) conditions
\begin{align}
& \textrm{tr}(\mathbf{Q}_z) \le  P_J, \mathbf{Q}_z \ge 0, \mathbf{g}_b \mathbf{Q}_z \mathbf{g}_b^H \le \varepsilon \\
& \lambda \ge 0, \mathbf{\Theta} \succeq 0 \\
& \textrm{tr}(\mathbf{\Theta} \mathbf{Q}_z) = 0 \label{trqphi} \\
& \lambda (\textrm{tr}(\mathbf{Q}_z)- P_J) = 0\\
& \mathbf{\Theta} = - \frac{\mathbf{g}_{e}^H \mathbf{g}_{e}}{(\mathbf{g}_{e} \mathbf{Q}_z \mathbf{g}_{e}^H)^2}  + \lambda \mathbf{I} + \nu \mathbf{g}_b \mathbf{g}_b^H.     \label{thetarank}
\end{align}
For the case that $\lambda = 0$, according to \eqref{thetarank}, we know that $\mathbf{\Theta}$ has a negative eigenvalue, which contradicts the fact that $\mathbf{\Theta}$ is positive semidefinite. Thus $\lambda$ can only be positive. For $\lambda > 0$, according to \eqref{thetarank} and since $\nu \ge 0$, $\mathbf{\Theta}$ has at least $N-1$ positive eigenvalues, \textit{i.e.} $\textrm{rank}(\mathbf{\Theta}) \ge N-1$.
\begin{lemma}[\cite{Marshall_Inequalities79}] \label{th:maj}
Given two $N \times N$ positive semidefinite matrices $\mathbf{A}$ and $\mathbf{B}$ with eigenvalues $\lambda_i(\mathbf{A})$ and $\lambda_i(\mathbf{B})$, respectively, arranged in non-increasing order, then
\begin{equation}
\textrm{tr}(\mathbf{AB}) \ge \sum_{i=1}^{N} \lambda_i(\mathbf{A}) \lambda_{N-i+1}(\mathbf{B}).
\end{equation}
\end{lemma}
Combining Lemma \ref{th:maj} and \eqref{trqphi}, assuming $\lambda_i(\mathbf{\mathbf{\Theta}})$ and $\lambda_{i}(\mathbf{Q}_z)$ are eigenvalues of $\mathbf{\mathbf{\Theta}}$ and $\mathbf{Q}_z$, respectively, in non-increasing order, we have
\begin{equation}
\textrm{tr}(\mathbf{\mathbf{\Theta}} \mathbf{Q}_z) = 0 \ge \sum_{i=1}^{N} \lambda_i(\mathbf{\mathbf{\Theta}}) \lambda_{N-i+1}(\mathbf{Q}_z).
\end{equation}
Due to the fact that $\mathbf{\mathbf{\Theta}}$ and $\mathbf{Q}_z$ are both positive semidefinite matrices, we have
\begin{equation}
\sum_{i=1}^{N} \lambda_i(\mathbf{\mathbf{\Theta}}) \lambda_{N-i+1}(\mathbf{Q}_z) = 0. \label{eq:phiq0}
\end{equation}
Thus we can conclude that $\textrm{rank}(\mathbf{\Theta}) \neq N$, since otherwise all eigenvalues of $\mathbf{Q}_z$ are zero and no jamming signals are transmitted. Combining this conclusion and the above observation that $\textrm{rank}(\mathbf{\Theta}) \ge N-1$ , we can conclude that $\textrm{rank}(\mathbf{\Theta}) = N-1$. Therefore, according to \eqref{eq:phiq0}, we have
\begin{equation}
\lambda_i(\mathbf{Q}_z)
\left \{ \begin{array}{l}
>0, \quad i=1 \\
=0, \quad i=2,\cdots,N,
\end{array}
\right.
\end{equation}
which indicates that $\textrm{rank}(\mathbf{Q}_z) = 1$, and the proof is complete.

\section{Proof of Proposition 3} \label{sec:maxqzpf}
The proof is similar to the proof of Proposition \ref{th:maxqx}. The maximin problem \eqref{eq:rjamor} can be transformed to
\begin{subequations} \label{eq:qzderi1}
\begin{align}
\max_{\mathbf{Q}_z \in \mathcal{Q}_z, v}& \quad v \\
{\textrm{s.t.}}& \quad (\tilde{\mathbf{g}}_e+\mathbf{e}_g) \mathbf{Q}_z (\tilde{\mathbf{g}}_e+\mathbf{e}_g)^H \ge v \\
					 & \quad\mathbf{e}_g \mathbf{e}_g^H \le \epsilon_g^2.
\end{align}
\end{subequations}
The constraints in \eqref{eq:qzderi1} can also be expressed as
\begin{subequations}
\begin{align}
&\mathbf{e}_g \mathbf{Q}_z \mathbf{e}_g^H + 2 {\textrm{Re}}(\tilde{\mathbf{g}}_e \mathbf{Q}_z \mathbf{e}_g^H)
+ \tilde{\mathbf{g}}_e \mathbf{Q}_z \tilde{\mathbf{g}}_e^H - v \ge 0 \\
&- \mathbf{e}_g \mathbf{e}_g^H + \epsilon_g^2 \ge 0.
\end{align}
\end{subequations}
According to the $\mathcal{S}$-procedure, there exists an $\mathbf{e}_g \in \mathbb{C}^{N_h}$ satisfying both of the above inequalities if and only if there exists a $\mu \ge 0$ such that
\begin{equation} \label{eq:appt1}
\left[
\begin{array}{cc}
\mu \mathbf{I}_{N_h} + \mathbf{Q}_z & \mathbf{Q}_z \tilde{\mathbf{g}}_e^H \\
\tilde{\mathbf{g}}_e \mathbf{Q}_z & \tilde{\mathbf{g}}_e \mathbf{Q}_z \tilde{\mathbf{g}}_e^H-\mu \epsilon_g^2 - v
\end{array}
\right] \succeq 0.
\end{equation}
Applying the Schur complement, we can rewrite \eqref{eq:appt1} as
\begin{equation}
 - \mu \epsilon_g^2 + \tilde{\mathbf{g}}_e \mathbf{Q}_z \tilde{\mathbf{g}}_e^H - \tilde{\mathbf{g}}_e \mathbf{Q}_z
(\mu \mathbf{I}_{N_h} + \mathbf{Q}_z)^\dag \mathbf{Q}_z \tilde{\mathbf{g}}_e^H \ge v.
\end{equation}
Therefore, the maximin problem \eqref{eq:rjamor} becomes
\begin{equation}
\max_{\mathbf{Q}_z \in \mathcal{Q}_z, \mu \ge 0} \quad
-\mu \epsilon_g^2 + \tilde{\mathbf{g}}_e \mathbf{Q}_z \tilde{\mathbf{g}}_e^H - \tilde{\mathbf{g}}_e \mathbf{Q}_z
(\mu \mathbf{I}_{N_h} + \mathbf{Q}_z)^\dag \mathbf{Q}_z \tilde{\mathbf{g}}_e^H,
\end{equation}
which is equivalent to
\begin{align}
\max_{\mathbf{Q}_z \in \mathcal{Q}_z, \mu \ge 0, \mathbf{\Psi}}& \quad -\mu \epsilon_g^2 + \tilde{\mathbf{g}}_e \mathbf{Q}_z \tilde{\mathbf{g}}_e^H - \tilde{\mathbf{g}}_e \mathbf{\Psi} \tilde{\mathbf{g}}_e^H \\
{\textrm{s.t.}}& \quad   \mathbf{Q}_z
(\mu \mathbf{I}_{N_h} + \mathbf{Q}_z)^\dag \mathbf{Q}_z \preceq \mathbf{\Psi}.
\end{align}
Next, we use the Schur complement to turn the above constraint into an LMI. The maximization problem is then given by
\begin{subequations}
\begin{align}
\max_{\mathbf{Q}_z, \mu, \mathbf{\Psi}}& \quad
-\mu \epsilon_g^2 + \textrm{tr} [(\mathbf{Q}_z-\mathbf{\Psi}) \tilde{\mathbf{g}}_e^H \tilde{\mathbf{g}}_e] \\
{\textrm{s.t.}}& \quad
\left[
\begin{array}{cc}
\mathbf{\Psi} & \mathbf{Q}_z \\
\mathbf{Q}_z & \mu \mathbf{I}_{N_h} + \mathbf{Q}_z
\end{array}
\right ] \succeq 0 \\
					 & \quad   \textrm{tr}(\mathbf{Q}_z) \le P_J \\
					 & \quad   \mathbf{Q}_z \succeq 0, \mu \ge 0,
\end{align}
\end{subequations}
and the proof is complete.

\section{Proof of Proposition 4} \label{sec:wegpf}
The proof is along the same line as that in Appendix \ref{sec:wehpf}. Problem \eqref{eq:maxeg} can be rewritten as
\begin{subequations}
\begin{align}
\min_{\mathbf{e}_g}& \quad  \mathbf{e}_g \mathbf{Q}_z^* \mathbf{e}_g^H + 2 \textrm{Re} (\tilde{\mathbf{g}}_{e} \mathbf{e}_g^H) + \tilde{\mathbf{g}}_{e} \mathbf{Q}_z^* \tilde{\mathbf{g}}_{e}^H \\
\textrm{s.t.}& \quad \mathbf{e}_g \mathbf{e}_g^H \le \epsilon_g^2,
\end{align}
\end{subequations}
with the Lagrangian
\begin{equation}
L(\mathbf{e}_g,\lambda)
= \mathbf{e}_g (\lambda \mathbf{I} + \mathbf{Q}_z^*) \mathbf{e}_g^H + 2 \textrm{Re} (\tilde{\mathbf{g}}_{e} \mathbf{Q}_z^* \mathbf{e}_g^H) + \tilde{\mathbf{g}}_{e} \mathbf{Q}_z^* \tilde{\mathbf{g}}_{e}^H  - \lambda \epsilon_g^2
\end{equation}
where $\lambda \ge 0$. It can be verified that the minimum of $L(\mathbf{e}_g,\lambda)$ with respect to $\mathbf{e}_g$ is achieved when $\mathbf{e}_g = -\tilde{\mathbf{g}}_e \mathbf{Q}_z^* (\lambda \mathbf{I} + \mathbf{Q}_z^*)^{-1}$, and hence the dual problem is
\begin{subequations}
\begin{align}
\max_{\lambda}& \quad \tilde{\mathbf{g}}_{e} \mathbf{Q}_z^* \tilde{\mathbf{g}}_{e}^H - \lambda \epsilon_g^2 -\tilde{\mathbf{g}}_{e} \mathbf{Q}_z^* (\lambda \mathbf{I} + \mathbf{Q}_z^*)^{-1} \mathbf{Q}_z^*  \tilde{\mathbf{g}}_{e}^H  \\
\textrm{s.t.}& \quad \lambda \mathbf{I} + \mathbf{Q}_z^* \succeq 0, \   \mathbf{Q}_z^* \tilde{\mathbf{g}}_{e}^H \in \mathcal{R}(\lambda \mathbf{I} + \mathbf{Q}_z^*).
\end{align}
\end{subequations}
Using a Schur complement, the dual problem becomes the following SDP
\begin{subequations}
\begin{align}
\max_{\lambda \ge 0, \gamma}& \quad \gamma \\
{\textrm{s.t.}}& \quad
\left[
\begin{array}{cc}
\lambda \mathbf{I} + \mathbf{Q}_z^* & \mathbf{Q}_z^* \tilde{\mathbf{g}}_{e}^H \\
\tilde{\mathbf{g}}_{e} \mathbf{Q}_z^* & \tilde{\mathbf{g}}_{e} \mathbf{Q}_z^* \tilde{\mathbf{g}}_{e}^H - \lambda \epsilon_g^2 - \gamma
\end{array}
\right ] \succeq 0,
\end{align}
\end{subequations}
which completes the proof.

\section{Proof of Proposition 5} \label{sec:appqosrcjp}
The maximin problem \eqref{eq:qosrcj} can be transformed to
\begin{subequations} \label{eq:qoseq}
\begin{align}
\min_{\mathbf{Q}_x \mathbf{Q}_z , v, t}& \quad \frac{v}{t} \\
{\textrm{s.t.}}& \quad (\tilde{\mathbf{h}}_e+\mathbf{e}_h) \mathbf{Q}_x (\tilde{\mathbf{h}}_e+\mathbf{e}_h)^H \le v \label{eq:sproceh1} \\
           & \quad \sigma^2+ (\tilde{\mathbf{g}}_e+\mathbf{e}_g) \mathbf{Q}_z (\tilde{\mathbf{g}}_e+\mathbf{e}_g)^H \ge t \label{eq:sproceg1}\\
		   & \quad \textrm{tr}(\mathbf{Q}_x) + \textrm{tr}(\mathbf{Q}_z) \le P \\
           & \quad |\mathbf{e}_h| \le \epsilon_h \label{eq:sproceh2}\\
           & \quad |\mathbf{e}_g| \le \epsilon_g \label{eq:sproceg2}\\
           & \quad \mathbf{h}_b \mathbf{Q}_x \mathbf{h}_b^H = \gamma_t(\mathbf{g}_b \mathbf{Q}_z \mathbf{g}_b^H + \sigma^2) .
\end{align}
\end{subequations}
Using the $\mathcal{S}$-procedure similarly as in \eqref{eq:sproc1}-\eqref{eq:sproc2} for \eqref{eq:sproceh1}, \eqref{eq:sproceh2} and for \eqref{eq:sproceg1}, \eqref{eq:sproceg2},
problem \eqref{eq:qoseq} can be rewritten as
\begin{equation}
\min_{\mathbf{Q}_x, \mathbf{Q}_z, \mu \ge 0, \nu \ge 0} \quad
\frac{\mu \epsilon_h^2 + \tilde{\mathbf{h}}_e \mathbf{Q}_x \tilde{\mathbf{h}}_e^H + \tilde{\mathbf{h}}_e \mathbf{Q}_x
(\mu \mathbf{I}_{N_a} - \mathbf{Q}_x)^\dag \mathbf{Q}_x \tilde{\mathbf{h}}_e^H}
{\sigma^2- \nu \epsilon_g^2 + \tilde{\mathbf{g}}_e \mathbf{Q}_z \tilde{\mathbf{g}}_e^H - \tilde{\mathbf{g}}_e \mathbf{Q}_z
(\nu \mathbf{I}_{N_h} + \mathbf{Q}_z)^\dag \mathbf{Q}_z \tilde{\mathbf{g}}_e^H},
\end{equation}
which is equivalent to
\begin{subequations}
\begin{align}
\min_{\mathbf{Q}_x, \mathbf{Q}_z, \mu \ge 0, \nu \ge 0, \mathbf{\Psi}, \mathbf{\Phi}}& \quad
\frac{\mu \epsilon_h^2 + \tilde{\mathbf{h}}_e \mathbf{Q}_x \tilde{\mathbf{h}}_e^H + \tilde{\mathbf{h}}_e \mathbf{\Psi} \tilde{\mathbf{h}}_e^H}
{\sigma^2- \nu \epsilon_g^2 + \tilde{\mathbf{g}}_e \mathbf{Q}_z \tilde{\mathbf{g}}_e^H - \tilde{\mathbf{g}}_e \mathbf{\Phi} \tilde{\mathbf{g}}_e^H} \\
{\textrm{s.t.}}& \quad   \mathbf{Q}_x
(\mu \mathbf{I}_{N_a} - \mathbf{Q}_x)^\dag \mathbf{Q}_x \preceq \mathbf{\Psi} \\
           & \quad   \mathbf{Q}_z
(\mu \mathbf{I}_{N_h} + \mathbf{Q}_z)^\dag \mathbf{Q}_z \preceq \mathbf{\Phi}.
\end{align}
\end{subequations}
Using the Schur complement for the above constraints, the expression in \eqref{eq:qosrcjp} can be obtained. 


\end{document}